# Discovering patterns of correlation and similarities in software project data with the Circos visualization tool

**Makrina Viola Kosti, Sofia Lazaridou, Nikoleta Bourazani, Lefteris Angelis**

Department of Informatics, Aristotle University of Thessaloniki 54124, Thessaloniki, GREECE
e-mail: {mkosti, solazari, nimpoura, lef,}@csd.auth.gr

**Abstract**

Software cost estimation based on multivariate data from completed projects requires the building of efficient models. These models essentially describe relations in the data, either on the basis of correlations between variables or of similarities between the projects. The continuous growth of the amount of data gathered and the need to perform preliminary analysis in order to discover patterns able to drive the building of reasonable models, leads the researchers towards intelligent and time-saving tools which can effectively describe data and their relationships. The goal of this paper is to suggest an innovative visualization tool, widely used in bioinformatics, which represents relations in data in an aesthetic and intelligent way. In order to illustrate the capabilities of the tool, we use a well known dataset from software engineering projects.

## 1 Introduction

Nowadays, data collection and accumulation, above different scientific fields, is growing dramatically. This leads to the need for efficient data preprocessing techniques and tools to help researchers in extracting useful information and make fast pre-modeling decisions. Various areas of informatics and computer science as well as many other scientific domains involving extensive data processing are living paradigms that face this need. Software cost estimation is such a field, since it is necessary for a researcher or a practitioner to look into and understand the relations governing the available data in order to build an efficient cost prediction model.

Data visualization refers to a wide collection of graphical techniques for data preprocessing and representation, which can lead to investigation and preliminary analysis of any multivariate dataset. There are several intelligent techniques which can be used to describe the relations within data while the intelligence concerns the smart way a multidimensional space is projected on two dimensions. These techniques can be used in several data-driven activities involving knowledge discovery, prediction, modeling (process/system), performing intelligent tasks (decision support, process planning, monitoring, diagnosis, etc.), pattern recognition etc.

Schroeder et al., (1998), underlining the increasing popularity of this approach, point out how complex scientific data can be displayed, where "complex" refers to multivariate datasets. Thus, raw experimental data are refashioned into an image or a series of images. Afterwards the image(s) can be explored. The latter is the essence of data visualization.

Regarding pattern extraction, Marchak (1994) notates the existence of a variety of techniques and explains how each one of these techniques can illustrate aspects of the information gathered that are not easily perceived by measures of central tendency or dispersion (also Loftus, 1993). Data visualization is part of the methods explained. McCormick et al. (1987) explain how scientific visualization involves computer graphics application and image processing techniques to provide awareness through visual methods by investigating "those mechanisms in humans and computers which allow them in concert to perceive use and communicate visual information".

One of the problems scientists have dealt with is related to the appropriate representation of multivariate datasets in multiple dimensions. In literature, a variety of approaches have been proposed, including Chernoff faces (Chernoff, 1973), harmonic function plots (Andrews, 1972) and three-dimensional box plots (Hartigan, 1975). These approaches were compared by Brown (1985). The findings showed that the methods varied in the ability to convey structure in the data and this was related to the familiarity with the representation method used each time. Du Toit et al. (1986) and Tukey (1977) also advocated the use of graphical representation in exploratory data analysis.

To this point we have made a brief review in relation with data visualization and graphical representation generally. We find it important, even though the literature does not provide clear directives on which method is more effective and which are the aspects differentiating those methods. From this point of view, a remarkable work has been done by statisticians. Specifically, Cleveland and McGill (1984) studied the effectiveness of different perceptual characteristics used to extract quantitative information from graphs. The study investigated length, direction, area and color amongst others. They rated the upper characteristics from the most accurate to the least accurate applying psychophysical methods. One year before, Grotch (1983) studied several methods to support interpretation of three-dimensional data visualizations.

Advances in computer technology, have contributed to more sophisticated graphical displays. In the context of scientific visualization, Castellan (1991) proposed that *"[powerful graphics] should enable scientists to better understand complex phenomena—particularly dynamic systems"*. In 2004 Martin Krzywinski presented Circos, a new visualization tool, originally designed for visualizing genomic data. A paragraph, optimally presenting this new emerging tool, could be (Krzywinski, M. et al., 2009, circos.ca): *"Circos attempts to bring a different aesthetic to science and strike a balance between flexibility and ease-of-use. Circos makes no assumptions about your data, uses extremely simple input data format, and makes image creation and customization easy. It's helping to make science look better, one figure at a time*."

We found that it would be interesting and helpful to apply this tool to software engineering data which are used for building cost prediction models. One reason for this is that Circos is not only applied to its originally designed type of biological datasets. As mentioned in Krzywinski, M. et al. (2009), Circos' circular layout makes it ideal for showing all types of data that include relationships between elements. Since in software cost estimation we are particularly interested in exploring correlations among variables for building statistical and probabilistic models (e.g. regression, Bayesian networks) and similarities between projects for case based reasoning approaches (e.g. estimation by analogy), Circos is a useful tool for data description, preliminary to model building.

In Section 2, we make a brief presentation of the tool. In Section 3, we analyze the application of the tool to a well known cost estimation dataset. The data used will be also presented in Section 3. Finally, in Section 4, we come to interpretation and some conclusions regarding the visualization of the chosen data.

## 2  Visual representation with CIRCOS

Circos is a mature software package that has been originally used, as we already mentioned, to mainly display genomic data (Constantine 2007; Jaillon et al. 2007; Campbell et al. 2008; Hampton et al. 2009; Meyer et al. 2009), also in online genomics resources (Forbes et al. 2008) and mainstream periodicals and newspapers (Constantine 2007; Duncan 2007; Ostrander 2007; Zimmer 2008). Additionally it is used to visualize data from other domains of study (Corum and Hossain 2007).

Circos uses a circular ideogram layout to facilitate the display of relationships between pairs of positions by the use of ribbons, which encode the position, size, and orientation of related genomic elements. It is capable of displaying data as scatter, line, and histogram plots, heat maps, tiles, connectors, and text. Bitmap or vector images can be created from GFF-style, (Perl extension for Gene Finding Feature format) data inputs and hierarchical configuration files, which can be easily generated by automated tools, making the tool suitable for rapid deployment in data analysis and reporting pipelines.

Primary, a configuration file is needed to give as input to the tool in order to perform visualization. Once the configuration file is ready, it is easy to hide elements, adjust scale (either globally or locally — a unique feature), or apply a different format to the data (visibility, opacity, shape, color and even position) based on dynamic rules.

The tool was designed to emphasize quantitative (i.e. similarity) or binary (i.e is or is not connected) relationships between data. Therefore it has many features that support the illustrating of the upper relations. Additionally its circular layout seems to be ideal in showing how different positions within data relate to one another. These relationships are depicted with colorful ribbons. You are allowed to change the thickness of the ribbons, representing the amount of relation, the progression and orientation of the circular segments and apply twists to the ribbons to indicate the orientation of the link. By doing all the above mentioned, one can easily and clearly show large amount of connections between data points. In Figure 1. we present a simple visualization example of a 4x4 distance matrix.. As we can observe, rows and columns are represented by circularly arranged segments (Row A, Row B, etc). The angular size of the segments is proportional to the total values of cells for the corresponding row or column. The values of the cells are represented by ribbons between a row and a column segment. However, a more detailed example is shown in the application of this tool using real project data.

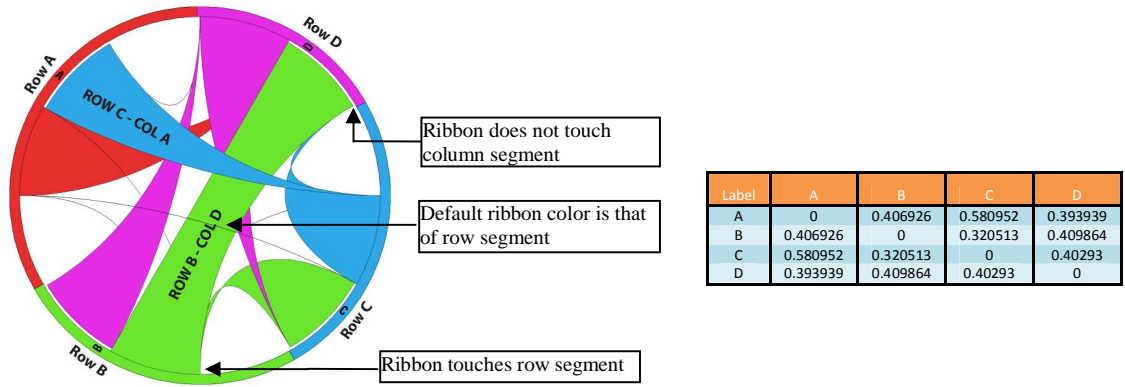

Figure 1. Visualization sample of a 4x4 matrix, using Circos

## 3 Application of CIRCOS

Case Based Reasoning and particularly the nearest neighbor method, has been used in a number of problem-solving areas. Among them is cost estimation, where the method is known as Estimation by Analogy (EbA). Conventional methods, despite the lack of a sound criterion for choosing nearest projects, were based on estimation using a fixed and predetermined number of neighbors from the entire set of historical instances. This approach puts boundaries to the estimation ability of such algorithms, for they do not take into consideration that every project under estimation is unique and requires different handling.

In this section we try to support this hypothesis by visualizing the pairwise similarities of the data. We do this using Circos, the visualization tool presented in Section 2. At this point we find it necessary to emphasize the objective of this work, which is not estimation, but exploration of data. This investigation of gathered information, as mentioned in Section 1, helps researcher to make decisions in the prediction domain.

In addition, we perform visualization of the correlation structure of data. Correlations between variables are especially useful for building and interpreting statistical and probabilistic models like least squares regression and Bayesian models. For example in regression analysis the correlations between the dependent variable and the independent variables are essential for selecting variables for a model and the correlations between independent variables are essential for avoiding the multicollinearity problem. Multicollinearity exists when there is strong correlation between two or more predictors in a regression model. Perfect collinearity exists when at least one predictor is a perfect linear combination of the others. In regression models is essential to avoid such phenomena which result in poor models with counter intuitive regression coefficients, difficult to be interpreted. A simple and initiatory way is by firstly visualizing these relations.

In Bayesian networks the correlations are represented by causal relationships. Furthermore, even in models functioning as "black boxes", like Neural Networks, it is essential to understand the relationships between variables. Consequently, our aim is to explore the possibility of using a powerful visualization tool in software cost estimation as an aid for the subsequent analysis and modeling procedure.

### 3.1 Data selection

In order to demonstrate our goal we decided to use the COCOMO NASA dataset, provided by the PROMISE DATA organization (http://promisedata.org/). The dataset is comprised of 60 NASA projects from different centers.

To calculate the similarities between the projects of the dataset, we chose a special dissimilarity coefficient suggested by Kaufman and Rousseeuw (1990). On the other hand, to extract correlations, we used the Spearman correlation coefficient, due to the large number of ordinal attributes of the data. The calculations were performed using Matlab R2009b.

### 3.2 Visualization - nearest analogy hypothesis

To support our hypothesis we generate a diagram using Circos visualization tool. It represents the pairwise similarities of projects with respect to the independent variables. The circular graph produced is shown in Figure 2. The visualization is further analyzed.

In Figure 2 we can see the pairwise distances of all projects of the dataset. The increasing thickness of ribbons depicts increasing similarities. The colors of the ribbons on the other hand show different ranges of similarities, starting with grey, which represents perfect feature similarity, then red, azure, mint and so on. Examining this diagram we can see that there is perfect similarity between projects 33 and 7, strong similarity between 32 and 28, quite strong similarity between 30 and 31, 26 and 27, 23 and 24 etc. We can also observe the following considerable neighborhoods (up to the mint ribbon):

- 28 is neighbor with 32, 31, 30, 29
- 9 is neighbor with 29, 28, 32, 10, etc
- 23 (24, 20, etc)

The neighborhoods created denote that estimating cost with a standard number of nearest neighbors for all projects is not realistic. We also can see that the graph is able to reveal groupings of projects, i.e. it can serve as a visual clustering method.

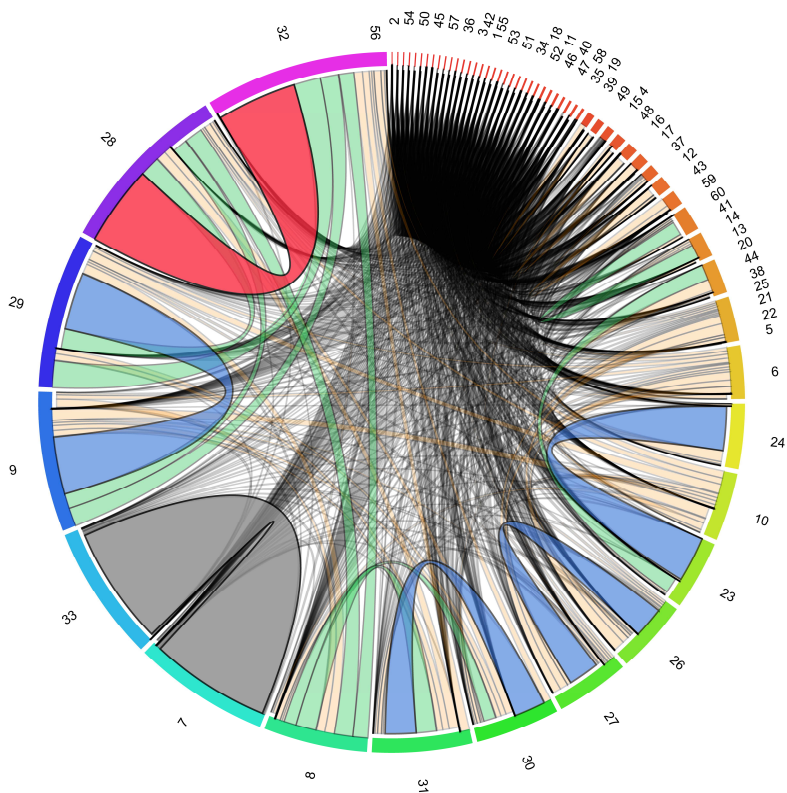

*Figure 2. Project pairwise distances for the COCOMO dataset*

We now present Figure 3 which represents the Spearman correlations of the ordinal and numerical attributes of the dataset.

We call up at this point that in regression analysis we do not want to have predictors that are strongly correlated. In Figure 2 the thickness of ribbons shows the correlation rate while the color represents the attribute the ribbon refers to. In this graph, black ribbons were made to depict negative correlations. According to this visualization there are some attributes that seem to be strongly correlated, i.e. ACAP (analysts capability) with AEXP (application experience) or STOR (main memory constraint) with TIME (time constraint for cpu), TIME (time constraint for cpu) with DATA (data base size) etc. This means that the data need further statistical analysis to avoid multicollinearity phenomena in the model.

Finally the correlation of the attributes with the effort can show how the dependent variable is affected by the independent ones. Specifically, the circos diagram reveals that effort (ACT_EFFORT) is positively correlated with LOC, DATA, TURN, ACAP, SCED etc variables while it is negatively correlated with CPLX, LEXP etc.

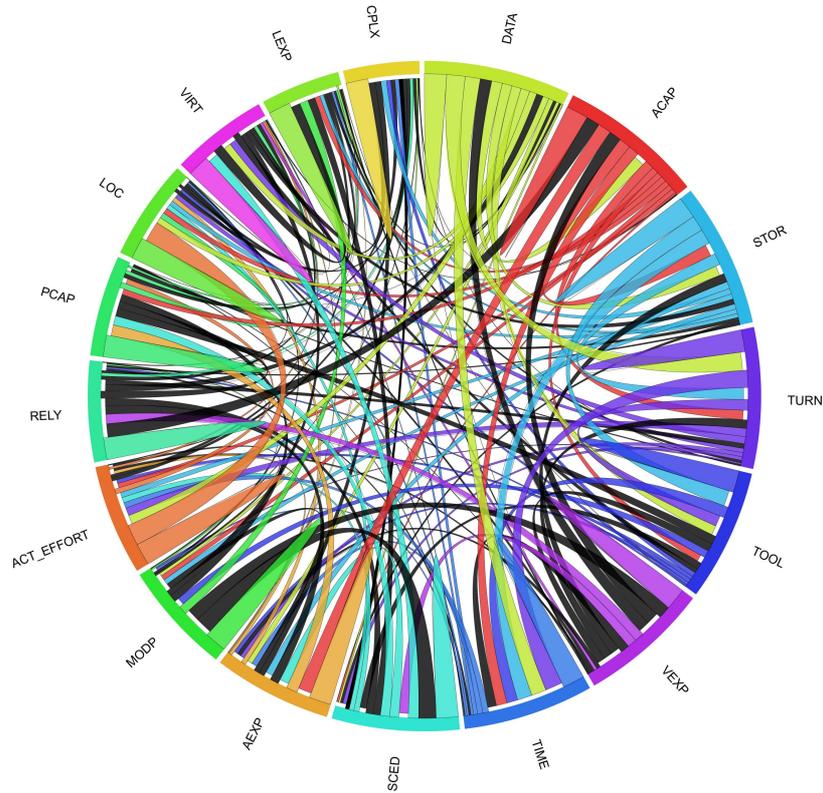

*Figure 3. Attribute correlations*

## 4 Conclusion

In this paper we present a visualization tool, originally designed by Martin Krzywinski for biological data. We firstly make a literature review to point out the benefits of data visualization as part of data descriptive processing. Data visualization via intelligent computer graphics makes simpler to understand the information that have been gathered, before proceeding to deeper analysis and modeling,

The new aesthetic visualization tool is called Circos. Circos applies a circular ideogram layout to facilitate the display of relationships between pairs of positions by the use of ribbons. We then try to demonstrate its abilities by visualizing similarities between software projects and correlations between variables of a widely known dataset, the COCOMO dataset. We chose similarities to demonstrate how the choice of a standard number of neighbors for the cost estimation of any project is not realistic. On the other hand we visualized correlations to point out how easily a visualized diagram can help us to detect possible multicollinearity phenomena between attributes.

In general, based on the experience of applying Circos to various project datasets, we can suggest the use of such tools, as they provide fast and easy way for displaying information which will be used in subsequent analysis and modeling.

# References

Andrews, D. F. (1972). Plots of high dimensional data. *Biometrics,* 28, 125-137.

Brown, R. L. (1985). Methods for the graphic representation of systems simulated data. *Ergonomics,* 28, 1439-1454.

Campbell PJ, Stephens PJ, Pleasance ED, O'Meara S, Li H, Santarius T, Stebbings LA, Leroy C, Edkins S, Hardy C, et al. 2008. Identification of somatically acquired rearrangements in cancer using genome-wide massively parallel paired-end sequencing. Nat Genet 40: 722–729.

Castellan, N. J., Jr. (1991).Computers and computing in psychology: Twenty years of progress and still a bright future. *Behavior Research Methods, Instruments, & Computers*, **23**, 106-108.

Chernoff, H. (1973). Usingfacesto representpointsin K-dimensional space graphically. *Journal of the American Statistical Association,*68, 361-368.

Cleveland, w. S., McGill, R. (1984). Graphicalperception:Theory, experimentationand application to the development of graphical methods. *Journal of the American Statistical Association, 79,* 531-554.

Constantine D. 2007. Close-ups of the genome, species by species by species. New York Times January 23, p. F4.

Corum J, Hossain F. 2007. Naming names. New York Times December 16, p. 41.

Duncan DE. 2007.Welcome to the future. Conde Nast Portfolio November: 192–197, 220–222.

Du Toit, S. H. C., Steyn, A. G. W., Stumpf, R. H. (1986). *Graphical exploratory data analysis.* New York: Springer-Verlag.

Forbes SA, Bhamra G, Bamford S, Dawson E, Kok C, Clements J, Menzies A, Teague JW, Futreal PA, Stratton MR. 2008. The Catalogue of Somatic Mutations in Cancer (COSMIC). Curr Protoc Hum Genet. 57: 10.11.1– 10.11.26. doi: 10.1002/0471142905.hg1011s57.

Grotch, S. L. (1983, November). Three-dimensional and stereoscopic graphs for scientific data display and analysis. *IEEE Computer Graphics & Applications,* 3(11), 31-43.

Hampton OA, Den Hollander P, Miller CA, Delgado DA, Li J, Coarfa C, Harris RA, Richards S, Scherer SE, Muzny DM, et al. 2009. A sequence-level map of chromosomal breakpoints in the MCF-7 breast cancer cell line yields insights into the evolution of a cancer genome. Genome Res 19: 167–177.

Hartigan, J. A. (1975). *Clustering algorithms.* New York: Wiley.

Jaillon O, Aury JM, Noel B, Policriti A, Clepet C, Casagrande A, Choisne N, Aubourg S, Vitulo N, Jubin C, et al. 2007. The grapevine genome sequence suggests ancestral hexaploidization in major angiosperm phyla. Nature 449: 463–467.

Kaufman, L., and Rousseeuw, P. J. 1990. "Finding Groups in Data: An Introduction to Cluster Analysis." New York: John Wiley.

Krzywinski, M. et al. Circos: an Information Aesthetic for Comparative Genomics. Genome Res (2009) 19:1639-1645

Loftus, G. (1993).A picture is worth a thousand *p* values: On the irrelevance of hypothesis testing in the microcomputer age. *Behavior Research Methods, Instruments, & Computers*, **25**, 250-256.

Marchak, F. M., & Whitney, D. A. (1990). Dynamic graphics in the exploratory analysis of multivariate data. *Behavior Research Methods, Instruments, & Computers*, **22**, 176-178.

McCormick, B. H., Defanti, T. A., Brown, M. D. (Eds.). (1987). Visualization in scientific computing [Special issue]. *Computer Graphics, 21(6).*

Meyer C, Kowarz E, Hofmann J, Renneville A, Zuna J, Trka J, Abdelali RB, Macintyre E, De Braekeleer E, De Braekeleer M, et al. 2009. New insights to the MLL recombinome of acute leukemias. Leukemia doi: 10.1038/ leu.2009.33.

Ostrander EA. 2007. Genetics and the shape of dogs. Am Sci 95: 406–413.

Schroeder,W., Martin, K., & Lorensen, B. (1998). *The Visualization Toolkit* (2nd ed.). Upper Saddle River, NJ: Prentice-Hall.

Tukey, J. W. (1977). *Exploratory data analysis.* Reading, MA: Addison-Wesley.

Zimmer C. 2008. Now: The rest of the genome. New York Times, November 11, p. D1.